\documentclass[twocolumn]{aastex631}

\usepackage{academicons}
\usepackage{amsmath}
\usepackage{bbm}
\usepackage{hyperref}
\usepackage{natbib}
\usepackage{xcolor}
\usepackage{xspace}
\usepackage{subfigure}
\usepackage{longtable}
\usepackage{booktabs}
\usepackage{threeparttable}

\newcommand{\Msini}{\ensuremath{M_c \sin i}\xspace}

\newcommand{\Mjup}{\ensuremath{M_{\mathrm{Jup}}}\xspace}

\usepackage{placeins}

\shortauthors{Van Zandt et al.}

\graphicspath{{./}{Figures/}}

\begin{document}

\title{A Smooth Transition from Giant Planets to Brown Dwarfs from the Radial Occurrence Distribution}

\correspondingauthor{Judah Van Zandt}
\email{judahvz@astro.ucla.edu}

\author[0000-0002-4290-6826]{Judah Van Zandt}
\altaffiliation{NASA FINESST Fellow}
\affiliation{Department of Physics \& Astronomy, University of California Los Angeles, Los Angeles, CA 90095, USA}

\author[0000-0003-0742-1660]{Gregory J. Gilbert}
\affiliation{Department of Physics \& Astronomy, University of California Los Angeles, Los Angeles, CA 90095, USA}
\affiliation{Department of Astronomy, California Institute of Technology, Pasadena, CA 91125, USA}

\author[0000-0003-0967-2893]{Erik A. Petigura}
\affiliation{Department of Physics \& Astronomy, University of California Los Angeles, Los Angeles, CA 90095, USA}

\author[0000-0002-8965-3969]{Steven Giacalone}
\altaffiliation{NSF Astronomy and Astrophysics Postdoctoral Fellow}
\affiliation{Department of Astronomy, California Institute of Technology, Pasadena, CA 91125, USA}

\author[0000-0001-8638-0320]{Andrew W. Howard}
\affiliation{Department of Astronomy, California Institute of Technology, Pasadena, CA 91125, USA}

\author[0000-0002-9305-5101]{Luke B. Handley}
\affiliation{Department of Astronomy, California Institute of Technology, Pasadena, CA 91125, USA}

\begin{abstract}
Measuring the occurrence rates of celestial objects is a valuable way to study their origins and evolution. Giant planets and brown dwarfs produce large Doppler signatures that are easily detectable by modern instrumentation, and legacy radial velocity (RV) surveys have now achieved full orbital coverage for periods $\lesssim$30 years. However, the Doppler method's sensitivity to companion minimum mass \Msini --- as opposed to true mass $M_c$ --- prevents unambiguous characterization using RVs alone, as purported giant planets may be brown dwarfs or stars on inclined orbits. Here we combined legacy RVs with absolute astrometry to re-fit the orbits of 194 companions from the California Legacy Survey. Around 40\% (7/18) of the ``brown dwarfs" (\Msini=13--80 \Mjup) we refit had true masses above 80 \Mjup. We incorporated our orbital posteriors and target sensitivity maps into a Poisson likelihood model to calculate occurrence as a function of true companion mass $M_c$ (0.8--80 \Mjup) and separation $a$ (0.3--30 AU). The semi-major axis distributions of objects in this range vary smoothly with mass, with Jupiter analogs favoring an abrupt increase in occurrence near 1 AU and brown dwarfs exhibiting a gradual enhancement at wider separations. Marginalized companion occurrence between 1--10 AU decreases smoothly with mass, with brown dwarfs having the lowest occurrence rate: $1.1^{+0.5}_{-0.4}\%$. Jupiter analogs are 10 times as common as brown dwarfs per mass interval in this range, demonstrating that the brown dwarf desert extends to 10 AU. The smooth variation in these distributions disfavors a sharp transition mass between ``bottom-up" core accretion and ``top-down" gravitational instability formation mechanisms, and rather suggests that these processes may produce companions in overlapping mass ranges.
\end{abstract}

\section{Introduction}
\label{sec:intro}

Understanding how giant planets and brown dwarfs form is a key question in exoplanet science. Planets below 1 \Mjup are thought to form ``bottom-up" through the collision of small planetesimals and subsequent runaway accretion of a gaseous envelope \citep{Pollack1996,YoudinZhu2025}. Meanwhile, objects above 80~\Mjup form ``top-down" by direct gravitational collapse of molecular clouds, and have sufficiently large masses to sustain core hydrogen burning \citep{McKeeOstriker2007, Shu1987}. Intermediate-mass objects (1--80 \Mjup) may occupy a region of overlap between these processes, with some larger objects forming through core accretion and smaller ones through direct collapse. Other formation channels could contribute in this regime as well, such as collapse due to gravitational instability of the protoplanetary disk (e.g. \citealt{KratterLodato2016}).

The distributions of intermediate-mass companion properties, such as their masses, orbital separations, and eccentricities, are useful tracers of their formation histories and may illuminate their relationship to the planetary and stellar populations. Companion occurrence near the ice line (1--10 AU) is particularly important due to the role that icy planetesimals play in planet formation \citep{DrazkowskaAlibert2017}. At present, detecting companions at separations near 10 AU is difficult. Direct imaging efforts have measured the prevalence of giant planets ($M_c$$\sim$5--13 \Mjup) and brown dwarfs ($M_c$$\sim$13--80 \Mjup) out to hundreds of AU, using evolutionary models (e.g. \citealt{Lagrange2025, Wilkinson2024, Baraffe03}) and recently, dynamical measurements incorporating absolute astrometry (e.g. \citealt{Franson2023, Currie2023}) to estimate these objects' masses. With a few notable exceptions, such as 51 Eridani b \citep{Macintosh2015}, $\beta$ Pictoris b \citep{Bonnefoy2014}, and AF Leporis b (\citealt{Franson2023, Mesa2023, DeRosa2023}), imaging has historically been unable to probe separations closer than $\sim$10 AU due to the high contrast between close companions and their host stars (see \citealt{BowlerNielsen2018} for a review). Meanwhile, few RV surveys have compiled the 30-year observing baselines needed to measure complete orbits of companions at 10 AU; these objects' signatures instead manifest as partial orbits which are difficult to distinguish from those caused by brown dwarfs and stellar companions.

One of the few RV samples that is sensitive out to 10 AU is the California Legacy Survey (CLS; \citealt{Rosenthal2021}). Using this sample's multi-decade RV data sets, \cite{Fulton2021} found that giant planet occurrence peaks between 1--10 AU and falls off at wider separations. The RV technique's sensitivity to companion minimum mass rather than true mass introduces ambiguity into the results of any Doppler survey. Given the high intrinsic prevalence of planetary objects \citep{Petigura2018, Rosenthal2022}, the probability of highly-inclined stars or brown dwarfs masquerading as planets is low. However, brown dwarfs are rare compared to stellar companions \citep{Raghavan2010}, resulting in a substantial probability that RV-discovered brown dwarfs are in fact inclined stars. Resolving the $\Msini$ ambiguity is therefore crucial to accurately measuring the occurrence of substellar companions.

In this study, we re-fit the RVs of 128 systems in the CLS, incorporating astrometric data from the \textit{Hipparcos}-\textit{Gaia} Catalog of Accelerations (HGCA; \citealt{Brandt2021}) when available, to constrain the inclinations and true masses of 194 stellar, substellar, and planetary companions (Sections~\ref{sec:previous_observations} and \ref{sec:analysis}). We present the occurrence of planets and brown dwarfs in this sample and note mass and semi-major-axis ranges with high and low formation efficiency (Section~\ref{sec:results}). Finally, we consider the implications of our measurements on formation of these objects (Section~\ref{sec:discussion}).

\section{Observations}
\label{sec:previous_observations}

\subsection{California Legacy Survey}
\label{subsec:cls}
The California Legacy Survey \citep[CLS;][]{Rosenthal2021} monitored the radial velocities (RVs) of 719 bright ($V\approx$ 6--10), nearby ($d\approx$ 8--60 pc) stars for more than 30 years. In total, this dataset includes 100,000 RVs from the  Keck/HIRES, APF/Levy, and Lick/Hamilton spectrographs. In this sample, \cite{Rosenthal2021} identified 226 companions with the following distribution of minimum masses: 34 stars ($\Msini > 80$~\Mjup), 19 RV brown dwarfs (\Msini = 13--80~\Mjup), and 173 RV planets\footnote{Throughout this article, we use the terms ``RV planet" and ``RV brown dwarf" to refer to objects with minimum masses $<13 \, \Mjup$ and 13--80 $\Mjup$, respectively.} ($\Msini < 13~\Mjup$). These detections spanned star-companion separations out to 20 AU and minimum masses $M_c \sin i$ down to ${\sim}0.01~\Mjup$. The well-defined selection function of the CLS, as well as the uniform orbital analysis performed by \cite{Rosenthal2021}, have made it a valuable resource for probing exoplanet demographics, particularly for planets at or near the water ice line \citep[e.g.][]{Fulton2021, VanZandtPetigura2024}.

CLS measured their system-by-system sensitivity to companions through injection/recovery tests. In brief, an injection/recovery test consists of randomly drawing a set of orbital parameters ($P$, $t_p$, $e$, $\omega$, $K$), modeling the RV signature that a companion with those parameters would produce, injecting that signature into a system's actual RV time series to capture system-specific RV noise and sampling inhomogeneities, and attempting to recover the signal with a planetary detection pipeline. CLS repeated this procedure 1000 times for each system, and estimated sensitivity as the fraction of successful recoveries in a given region of $M_c$-$a$ space. We calculated CLS's survey-averaged sensitivity by combining the injection/recovery tests of all stars in the sample. We show this average map, together with the CLS catalog, in both panels of Figure \ref{fig:catalog_multipanel}.

\subsection{Hipparcos-Gaia Catalog of Accelerations}
\label{subsec:hgca}

The $Hipparcos$-$Gaia$ Catalog of Accelerations (HGCA; \citealt{Brandt2021}) provides astrometric proper motion measurements for over 115,000 stars calculated by aligning the \textit{Hipparcos} reference frame \citep{HipparcosCatalog} with that of $Gaia$ EDR3 \citep{GaiaEDR3}. The HGCA lists three separate measurements of stellar proper motion: the $Hipparcos$ and $Gaia$ epoch proper motions, measured at approximately 1991.25 and 2016.0, respectively, as well as an average proper motion computed from the difference in the stellar position at these two epochs. By identifying changes in proper motion over time, one may infer the presence of massive long-period companions. The HGCA has found particular success in concert with other detection techniques, either by highlighting promising systems for follow-up observation (e.g., \citealt{Franson2023}) or by improving measurement precision through joint fitting (e.g., \citealt{VanZandt2024, Zhang2025}). Here we use the second approach, combining HGCA accelerations with the multi-decade RV baselines of the CLS to fit orbits in three dimensions.

\section{Analysis}\label{sec:analysis}

\subsection{Joint RV/astrometry orbital fits}
\label{subsec:orbit_fits}

We refit the orbits of 191 companions in the CLS with $\Msini < 80$ \Mjup, plus three objects with $\Msini > 80$ \Mjup in systems with substellar companions, for a total of 194 orbital fits among 128 unique systems. We used \texttt{Orvara} \citep{Orvara2021} to fit three-dimensional orbits, parameterized by primary mass $M_{\text{pri}}$, companion mass $M_{c}$, semi-major axis $a$, eccentricity $e$, inclination $i$, argument of periastron of the companion orbit $\omega$, longitude of ascending node $\Omega$, parallax $\varpi$, and mean longitude at a reference epoch $\lambda_{\text{ref}}$. Fitting three-dimensional orbits allowed us to constrain companion true mass rather than \Msini. \texttt{Orvara} supports joint fits of RV time series and absolute astrometry from both $Hipparcos$ and $Gaia$, including modeling of the $Hipparcos$ intermediate astrometric data and $Gaia$ scanning law, through its integration of the \texttt{htof} astrometric fitting package \citep{htof2021}.

We used the RV measurements provided in \cite{Rosenthal2021} for all systems. We supplemented these RVs with literature measurements, incorporating velocities from the ELODIE \citep{Baranne1996}, UCLES \citep{Diego1996}, HRS \citep{Tull1998}, HIDES \citep{Izumiura1999}, CORALIE \citep{Queloz2000}, HARPS \citep{Pepe2000}, MIKE \citep{Bernstein2003}, SOPHIE \citep{Perruchot2008}, and NEID \citep{Schwab2016} spectrographs. For each system, \texttt{Orvara} analytically marginalizes over the systemic RV zero point, and assumes one RV jitter term which is common to all instruments. HIRES was upgraded in August 2004 \citep{Butler2006}, and \cite{Rosenthal2021} presents both pre- and post-2004 RVs. We fit the offset associated with this upgrade, treating HIRES RVs collected before/after August 2004 as originating from different instruments. Three of the other instruments listed above also underwent upgrades documented in the literature: HARPS in 2015 \citep{LoCurto2015}, SOPHIE in 2011 \citep{Perruchot2011}, and CORALIE in 2007 and 2014 (see \citealt{Rickman2019} and references therein). In the relatively low number of systems in which we used RVs from these instruments (we used HARPS RVs in 3 systems, SOPHIE RVs in 13 systems, and CORALIE RVs in 14 systems), all RVs were collected prior to the instrument upgrade(s). Thus, no zero point offset corrections were necessary for these RVs.
 
\begin{figure*}[t]
    \centering
    \begin{minipage}[b]{0.8\linewidth}
        \centering
        \includegraphics[width=\linewidth]{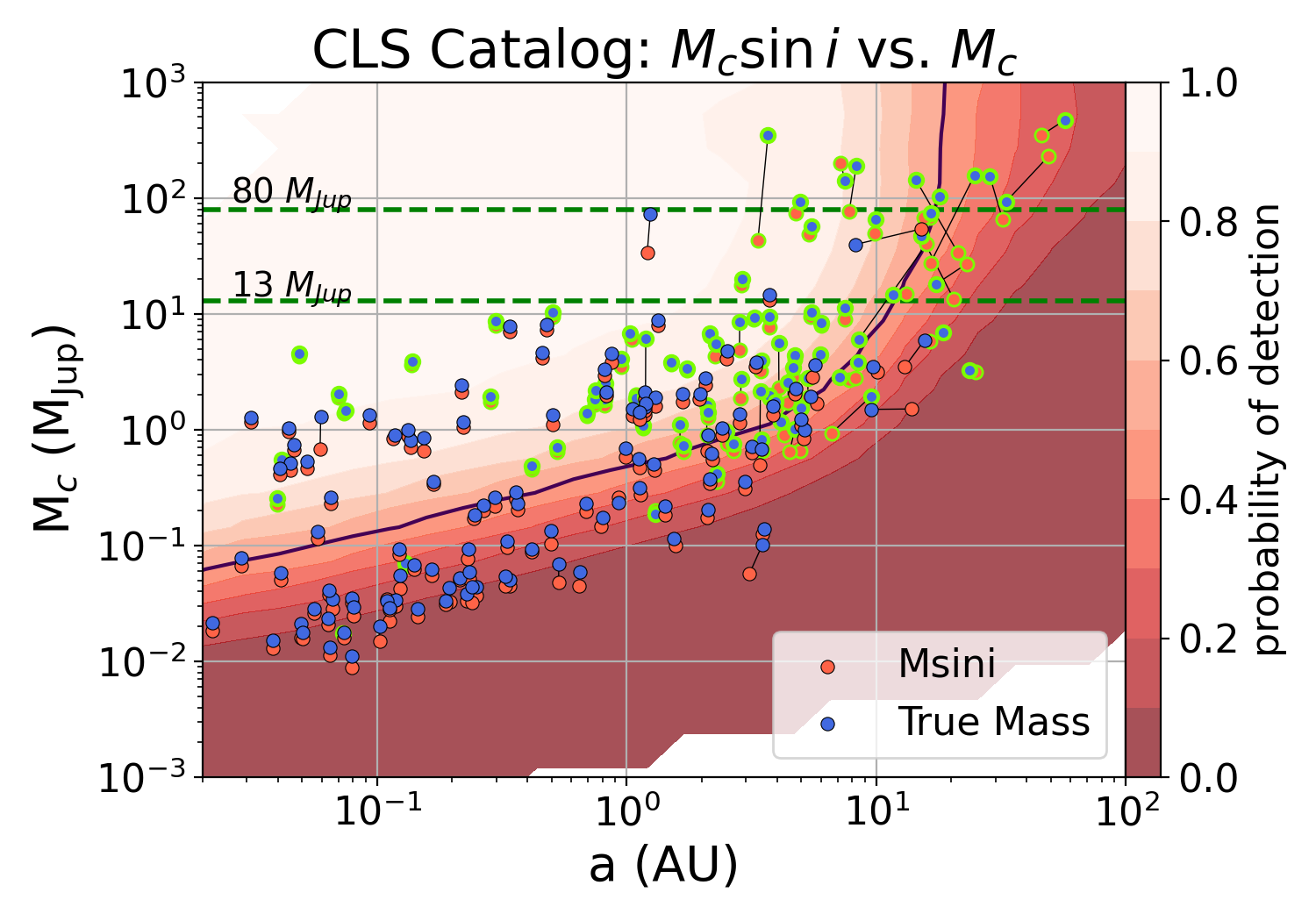}
    \end{minipage}
    \hfill
    \begin{minipage}[b]{0.8\linewidth}
        \centering
        \includegraphics[width=\linewidth]{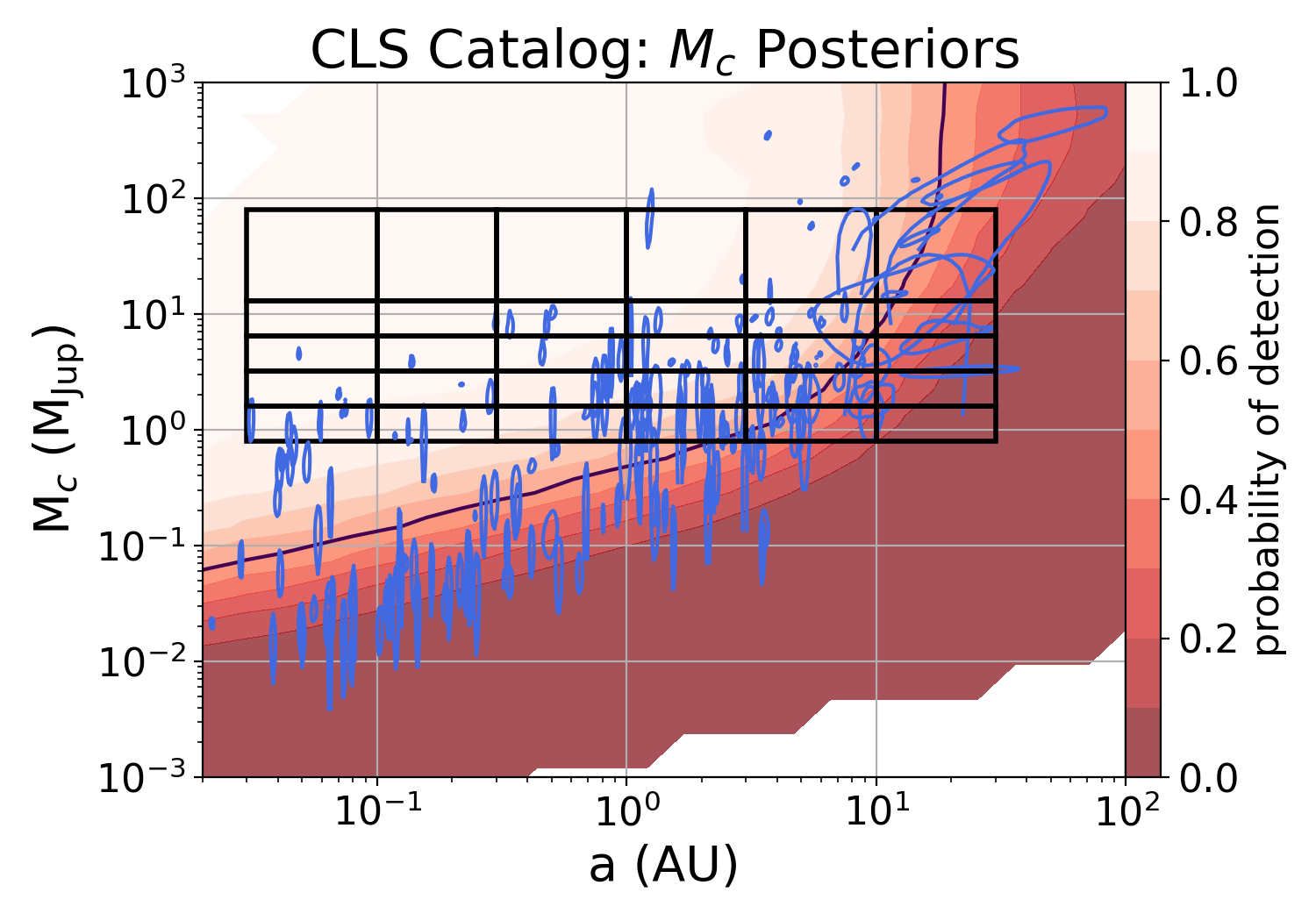}
    \end{minipage}
    
    \caption{\textbf{Top:} The masses and semi-major axes of the substellar ($\Msini \leq 80 \, \Mjup$) companions reported in the CLS catalog. Orange points show the average $a$ and \Msini fitted, or in some cases quoted, by CLS, while blue points show the average $a$ and $M_c$ refitted in this work. Green borders indicate companions in whose orbital fits we included HGCA astrometry. The color map shows the CLS sensitivity to true mass, averaged over all 719 stars in the survey, with contours indicating sensitivity deciles and the 50\% completeness line marked in black. Horizontal green lines mark 13 and 80 \Mjup, the traditional brown dwarf mass limits. \textbf{Bottom:} Our re-fit catalog of 194 CLS companions. Contours enclose 68\% of an individual companion's posterior draws. Black rectangles indicate the cells in which we calculated companion occurrence (see Section \ref{sec:results}). Our occurrence results flow from the companions whose posteriors fall fully or partially within these rectangles. The completeness map is the same as in the top panel. Note that because the colored markers in the top panel indicate averages, companions with $M_c>80 \, \Mjup$ may still have significant posterior mass below this limit, and vice versa. We highlight some of the outliers in this plot in Figure \ref{fig:catalog_appendix}.}
    \label{fig:catalog_multipanel}
\end{figure*}

We included astrometric accelerations from the HGCA for the 53 systems in which the acceleration was significant ($\frac{\Delta\mu}{\sigma_{\Delta\mu}}\geq3$). In the remaining 75 systems, we supplied no astrometric data and fit only the RVs, allowing \texttt{Orvara} to marginalize over our uncertainty in inclination. The companions in systems with significant astrometric accelerations are indicated in the top panel of Figure \ref{fig:catalog_multipanel}. As expected, accelerating systems tend to host more massive companions: all but four systems exhibiting significant astrometric acceleration host at least one companion with $M_c > 1$ \Mjup.

We conducted orbital fits in stages, successively tuning our models for systems for which our MCMC simulations did not converge. For each system, we began with a model informed by the results of R21. We assumed the same number of companions they found, and bounded the range of companion eccentricity and semi-major axis at $\pm4\sigma$ of their fitted values. In systems with significant trends, we included an additional companion to the model with a minimum separation of 10 AU. We widened the prior limits on eccentricity and/or semi-major axis for companions whose orbital posteriors were non-zero at the bounds. We applied \texttt{Orvara}'s default priors, including log-uniform priors on mass and semi-major axis. We applied this bounding procedure based on R21 in conjunction with standard priors merely to prevent the sampler from wandering into very low probability regions of parameter space where MCMC chains were liable to become stuck; our use of R21 posteriors to inform the model initialization thus does not double-condition the data by introducing an erroneous prior. \texttt{Orvara} implements a parallel-tempering MCMC sampler called \texttt{ptemcee} \citep{Vousden2016}, itself a fork of \texttt{emcee} \citep{ForemanMackey2013}, which allows simultaneous sampling from separate versions of the posterior, each smoothed according to a ``temperature" parameter. Our default MCMC parameters for each system were as follows: we used 100 walkers, 25 temperatures, and ran each simulation for 200,000 steps with a 100,000-step burn-in.  We required that each burned-in chain be at least 50 times the autocorrelation length, and iteratively increased the number of steps for systems that failed this criterion.

We deviated from the above prescription for a handful of systems that required custom treatment. CLS reported two companions to \textbf{HD 167215}, one stellar and one substellar, at 7.2 and 8 AU, respectively. We found that the longer-period signal was likely a residual caused by an imperfect fit to the inner stellar companion orbit, and included only the latter in our model. CLS also reported an activity-induced false positive signal in the \textbf{HD 192310} system. We quoted their one-companion model parameters directly and marginalized over the companion's unknown inclination by dividing each \Msini sample by $\sin i$, with $i$ drawn from a $\cos i$-uniform distribution. We did the same for \textbf{HD 219134}, for which CLS reported five planets and three false positive signals associated with annual or instrumental systematics. We quoted the parameters of the two companions to \textbf{HD 28185} from \cite{Venner2024}, who incorporated 39 extra RVs and HGCA astrometry into their analysis. Notably, they revised the mass of HD 28185 c to a planetary value ($M_c=6.0 \pm 0.6 \, \Mjup$), whereas CLS reported a minimum mass range consistent with stellar, brown dwarf, and planetary companions ($M_c\sin i=40^{+40}_{-30} \, \Mjup$). Finally, we quoted CLS's orbital parameters and an inclination of 59.5$^{\circ}$ for all three planets \citep{Rivera2010} for the resonant system \textbf{GL 876}. \cite{Rivera2010} reported four planets in this system, but CLS found only three after ruling out multiple spurious signals. In a further eight systems, we included an additional long-period companion to model non-periodic RV variability. Given the large uncertainties on the fitted parameters of these companions, we omitted them from our occurrence calculations.


After removing the spurious substellar companion to HD 167215, we were left with 18 RV brown dwarfs (\Msini = 13--80~\Mjup). Of these, seven had fitted true masses greater than 80 \Mjup. We compare the original CLS catalog with our refit parameters in the left panel of Figure \ref{fig:catalog_multipanel}, and show the refit posteriors in the right panel. The original and refit parameters of the RV brown dwarf systems are highlighted in Table \ref{tab:bd_before_after}, and refits for the full sample are given in Appendix \ref{app:all_before_after_table}.

\begin{deluxetable*}{lcccccccccc}
\tablecaption{Original and Refit Parameters for Companions with $M_c \sin i \, = 13$--$\, 80 \, \Mjup$}
\label{tab:bd_before_after}
\tablehead{
\colhead{CLS} &
\colhead{$a_i$} &
\colhead{$M_c \sin i$} &
\colhead{$e_i$} &
\colhead{$a_f$} &
\colhead{$M_{c}$} &
\colhead{$e_f$} &
\colhead{$\omega_f$} &
\colhead{$i_f$} &
\colhead{$\Omega_f$} &
\colhead{}\\
\colhead{Name} &
\colhead{(AU)} &
\colhead{($\Mjup$)} &
\colhead{} &
\colhead{(AU)} &
\colhead{($\Mjup$)} &
\colhead{} &
\colhead{($^{\circ}$)} &
\colhead{($^{\circ}$)} &
\colhead{($^{\circ}$)} &
\colhead{HGCA?}
}
\startdata
  HD 111031 &       $30^{+20}_{-10}$ &     $70^{+50}_{-30}$ &       $0.5^{+0.1}_{-0.2}$ &           $28^{+11}_{-6}$ &    $160^{+40}_{-30}$ &       $0.5^{+0.1}_{-0.1}$ &      $110^{+20}_{-20}$ &          $156^{+4}_{-6}$ &      $347^{+6}_{-6}$ &        True \\
  HD 126614 &         $16^{+2}_{-2}$ &       $28^{+6}_{-5}$ &    $0.04^{+0.05}_{-0.03}$ &           $25^{+18}_{-8}$ &   $160^{+220}_{-60}$ &       $0.1^{+0.2}_{-0.1}$ &       $30^{+60}_{-40}$ &          $28^{+58}_{-9}$ &     $80^{+20}_{-20}$ &        True \\
   HD 16160 &   $16.4^{+0.3}_{-0.3}$ &       $67^{+2}_{-2}$ & $0.643^{+0.004}_{-0.004}$ &      $17.9^{+0.2}_{-0.2}$ &      $102^{+3}_{-3}$ & $0.658^{+0.002}_{-0.002}$ &  $133.5^{+0.1}_{-0.1}$ &     $46.0^{+0.7}_{-0.7}$ &      $306^{+1}_{-1}$ &        True \\
  HD 167665 & $5.39^{+0.08}_{-0.08}$ &       $48^{+1}_{-1}$ & $0.342^{+0.004}_{-0.004}$ &    $5.48^{+0.08}_{-0.08}$ &       $57^{+3}_{-2}$ & $0.343^{+0.003}_{-0.003}$ & $-135.8^{+0.5}_{-0.5}$ &         $114^{+4}_{-56}$ &       $38^{+1}_{-1}$ &        True \\
HD 168443 c & $2.88^{+0.03}_{-0.03}$ & $17.8^{+0.4}_{-0.4}$ & $0.211^{+0.001}_{-0.001}$ &    $2.91^{+0.03}_{-0.02}$ &       $20^{+1}_{-1}$ & $0.214^{+0.002}_{-0.002}$ &   $66.7^{+0.5}_{-0.5}$ &         $70^{+40}_{-10}$ &     $300^{+11}_{-5}$ &        True \\
   HD 18445 & $1.21^{+0.02}_{-0.02}$ &       $34^{+7}_{-4}$ &    $0.67^{+0.12}_{-0.10}$ &    $1.24^{+0.02}_{-0.02}$ &     $70^{+50}_{-10}$ &    $0.89^{+0.05}_{-0.06}$ &        $98^{+10}_{-6}$ &        $100^{+40}_{-50}$ &  $200^{+100}_{-100}$ &       False \\
  HD 190406 &   $15.5^{+0.3}_{-0.3}$ &       $67^{+2}_{-2}$ & $0.462^{+0.005}_{-0.005}$ &      $16.5^{+0.3}_{-0.3}$ &       $73^{+2}_{-2}$ & $0.465^{+0.004}_{-0.004}$ &  $-93.3^{+0.6}_{-0.6}$ &           $98^{+2}_{-2}$ &      $331^{+1}_{-2}$ &        True \\
  HD 211681 &    $7.8^{+0.2}_{-0.2}$ &       $76^{+3}_{-3}$ & $0.441^{+0.008}_{-0.006}$ &       $8.3^{+0.2}_{-0.2}$ &      $189^{+8}_{-8}$ & $0.449^{+0.003}_{-0.003}$ &  $-53.6^{+0.6}_{-0.6}$ &    $153.2^{+0.5}_{-0.5}$ &      $208^{+1}_{-1}$ &        True \\
  HD 239960 &         $15^{+7}_{-5}$ &     $50^{+10}_{-10}$ &       $0.6^{+0.1}_{-0.1}$ &       $8.2^{+1.0}_{-0.6}$ &      $40^{+23}_{-6}$ &    $0.36^{+0.05}_{-0.04}$ &       $-150^{+4}_{-4}$ &         $90^{+40}_{-40}$ &  $200^{+100}_{-100}$ &       False \\
 HD 26161 b &         $20^{+8}_{-5}$ &       $14^{+8}_{-4}$ &    $0.82^{+0.06}_{-0.05}$ &            $15^{+2}_{-1}$ &       $47^{+7}_{-5}$ &    $0.84^{+0.02}_{-0.01}$ &   $-5.7^{+0.8}_{-0.9}$ &         $126^{+9}_{-76}$ &      $29^{+5}_{-13}$ &        True \\
   HD 28185 &         $16^{+7}_{-5}$ &     $40^{+40}_{-30}$ &    $0.26^{+0.12}_{-0.09}$ &    $8.50^{+0.29}_{-0.26}$ &  $6.0^{+0.6}_{-0.6}$ &    $0.15^{+0.04}_{-0.04}$ &                    --- &                      --- &                  --- &        True \\
 HD 38529 b & $3.74^{+0.01}_{-0.01}$ & $13.2^{+0.1}_{-0.1}$ & $0.354^{+0.005}_{-0.005}$ & $3.742^{+0.010}_{-0.010}$ &       $15^{+4}_{-1}$ & $0.354^{+0.005}_{-0.005}$ &         $21^{+1}_{-1}$ &         $90^{+30}_{-30}$ &  $200^{+100}_{-100}$ &       False \\
    HD 4747 &    $9.8^{+0.2}_{-0.2}$ &       $49^{+2}_{-2}$ & $0.731^{+0.002}_{-0.002}$ &       $9.9^{+0.1}_{-0.1}$ &       $66^{+2}_{-2}$ & $0.730^{+0.001}_{-0.001}$ &  $-92.6^{+0.5}_{-0.5}$ &           $50^{+1}_{-1}$ &       $91^{+1}_{-1}$ &        True \\
 HD 66428 c &        $23^{+19}_{-8}$ &     $30^{+20}_{-20}$ &       $0.3^{+0.2}_{-0.2}$ &           $17^{+11}_{-6}$ &      $18^{+11}_{-6}$ &    $0.21^{+0.20}_{-0.07}$ &     $-120^{+40}_{-30}$ &        $130^{+30}_{-60}$ &    $140^{+20}_{-20}$ &        True \\
   HD 68017 &         $21^{+5}_{-4}$ &       $34^{+6}_{-6}$ &    $0.43^{+0.08}_{-0.09}$ &      $14.4^{+0.3}_{-0.3}$ &      $143^{+2}_{-2}$ & $0.296^{+0.010}_{-0.010}$ &        $-75^{+2}_{-2}$ & $170.00^{+0.07}_{-0.08}$ & $97.7^{+0.5}_{-0.5}$ &        True \\
 HD 68988 c &         $13^{+5}_{-2}$ &       $15^{+3}_{-2}$ &    $0.45^{+0.13}_{-0.08}$ &      $11.7^{+0.9}_{-0.7}$ & $14.6^{+0.8}_{-0.6}$ &    $0.37^{+0.04}_{-0.04}$ &        $173^{+2}_{-3}$ &         $80^{+10}_{-20}$ &    $350^{+10}_{-10}$ &        True \\
    HD 8765 & $3.36^{+0.05}_{-0.05}$ &       $43^{+2}_{-2}$ & $0.394^{+0.010}_{-0.010}$ &    $3.68^{+0.05}_{-0.05}$ &    $350^{+20}_{-20}$ & $0.510^{+0.006}_{-0.006}$ &          $8^{+1}_{-1}$ &     $10.5^{+0.4}_{-0.4}$ &      $198^{+2}_{-2}$ &        True \\
  HIP 63510 & $4.78^{+0.06}_{-0.05}$ &       $74^{+3}_{-3}$ &    $0.31^{+0.03}_{-0.02}$ &    $4.95^{+0.04}_{-0.04}$ &       $93^{+2}_{-2}$ &    $0.30^{+0.03}_{-0.02}$ &        $149^{+6}_{-6}$ &          $128^{+2}_{-2}$ &       $57^{+3}_{-3}$ &        True \\
\enddata
\tablenotetext{}{%
\textbf{Note.} Initial values, indicated with a subscript $i$, as well as $M_c \sin i$ measurements, are from \cite{Rosenthal2021}. Final values, indicated with a subscript $f$, and true mass measurements are from this work. The one exception is HD 28185, whose refit parameters we quote from \cite{Venner2024}. The left-most column gives companion names as they appear in \cite{Rosenthal2021}. The right-most column indicates systems for which we included the HGCA acceleration in the fit.
}
\end{deluxetable*}

\subsection{Occurrence Framework}
\label{subsec:occurrence_framework}

We modeled companion occurrence as a piecewise constant function, that is, a histogram, in companion mass $M_c$ and semi-major axis $a$. We assumed that the companion occurrence rate in each histogram bin was independent of those in other bins.

Following previous studies (e.g., \citealt{ForemanMackey2014, Fulton2021, Rogers2021, VanZandt2025}), we modeled our companion catalog as a series of draws from a censored Poisson point process defined by the companion occurrence rate density (ORD) $\lambda(\log a, \log M_c|\boldsymbol{\theta})$, which gives the number of companions per star per logarithmic $a$-$M_c$ interval. We define $\boldsymbol{\omega} = (\log a, \, \log M_c)$ for notational compactness and use $\boldsymbol{\theta}$ to denote the vector of model parameters of $\lambda$, which for a histogram model is simply the bin heights.

The hierarchical model likelihood comprises two terms: the probability of detecting $K$ companions with the observed $a$ and $M_c$, and the probability of detecting no more companions in the rest of the parameter space. The likelihood for the $j^{\text{th}}$ histogram bin, i.e., two-dimensional interval in $\log a$---$\log M_c$ space, is

\begin{equation}
    \mathcal{L}_j \propto e^{-\Lambda_j} 
    \prod_{k=1}^{K} 
    \Bigg [
    \frac{1}{N_{k}}
    \sum_{n=1}^{N_{k}} \frac{\lambda(\boldsymbol{\omega}_{k}^{(n)}|\boldsymbol{\theta})Q(\boldsymbol{\omega}_{k}^{(n)})}{p(\boldsymbol{\omega}_{k}^{(n)}|\boldsymbol{\alpha})} \cdot \mathbbm{1}_{n,k,j}
    \Bigg ]^{w_{k,j}}.
    \label{eq:likelihood_single_bin}
\end{equation}

\noindent In the first term, the Poisson rate parameter $\Lambda_j$ gives the expected total number of observed companions among $N_\star$ host stars in the $j^{\text{th}}$ bin:

\begin{equation}
    \Lambda_j = N_* \int \lambda(\boldsymbol{\omega}|\boldsymbol{\theta}) \, Q(\boldsymbol{\omega}) \, d\boldsymbol{\omega},
    \label{eq:Lambda_definition}
\end{equation}

\noindent where the integral is over the $j^{\text{th}}$ bin and $Q$, the average sensitivity of the CLS to companions as a function of $a$ and $M_c$, accounts for missed companions.

The second term gives the probability of detecting $K$ companions with the observed parameters $\{ \boldsymbol{\omega}_k^{(n)} \}$:

\begin{equation}
    P(\boldsymbol{\omega} | \boldsymbol{\theta}) = 
    \prod_{k=1}^{K} 
    \Bigg [
    \frac{1}{N_{k}}
    \sum_{n=1}^{N_{k}} \frac{\lambda(\boldsymbol{\omega}_{k}^{(n)}|\boldsymbol{\theta})Q(\boldsymbol{\omega}_{k}^{(n)})}{p(\boldsymbol{\omega}_{k}^{(n)}|\boldsymbol{\alpha})} \cdot \mathbbm{1}_{n,k,j}
    \Bigg ]^{w_{k,j}}.
    \label{eq:likelihood_product_term}
\end{equation}

We use $k$ and $n$ to index the companions and posterior samples, respectively, so that $\boldsymbol{\omega}_{k}^{(n)}$ represents the $n^{\text{th}}$ draw from the posterior of companion $k$. We included $K=191$ companions in our occurrence analysis, and sampled $N_k=1000$ draws from each posterior. The indicator function $\mathbbm{1}_{n,k,j}$ equals one when $\boldsymbol{\omega}_{k}^{(n)}$ falls in the $j^{\text{th}}$ bin, and zero otherwise. The interim prior $p$ is the prior we used during orbit fitting: log-uniform in $a$ and $M_c$.

Many of the posteriors in our catalog have fractional uncertainties that are comparable to the bin sizes. Following \cite{Gilbert2025}, we accounted for these large errors by weighting each companion's contribution to the likelihood. For each companion/bin pair, we calculated a weight, $w_{k,j}$, equal to the fraction of draws from that companion's posterior that fell within the bin.

The full likelihood over $J$ bins is

\begin{equation}
    \mathcal{L} \propto
    \prod_{j=1}^{J} 
    \mathcal{L}_j.
    \label{eq:likelihood_full}
\end{equation}

\noindent We placed uniform priors on the bin heights $\boldsymbol{\theta}$ and sampled our likelihood using \texttt{emcee}.

We repeated our fitting procedure with the extra requirement that the histogram bin heights vary smoothly. We enforced this assumption by drawing bin heights from a Gaussian Process (GP) prior defined by a two-dimensional Mat\'ern--3/2 kernel, with covariance function

\begin{equation}
    K(\omega_i, \omega_j) = \sigma^2 
    \left ( 1 + \sqrt{3}d \right ) \text{exp}\left ( -\sqrt{3}d \right ),
    \label{eq:matern_kernel}
\end{equation}

\noindent where $d^2 = (\frac{\Delta \log a}{\ell_a})^2 + (\frac{\Delta \log M_c}{\ell_{M_c}})^2$ and the covariance scale $\sigma$, the semi-major axis correlation length $\ell_a$, and the mass correlation length $\ell_{M_c}$ are hyperparameters of the model. We tested both log-normal and log-uniform hyperpriors on these parameters, finding in both cases that the parameters were essentially unconstrained by the data. Given the statistically negligible effect of GP smoothing on our results, we report our non-smoothed results in the remainder of this work.

\section{Results}
\label{sec:results}

\subsection{Companion occurrence as a function of mass and semi-major axis}
\label{subsec:mass_sma_occurrence}

\begin{figure}
    \centering
    \includegraphics[width=0.90\linewidth]{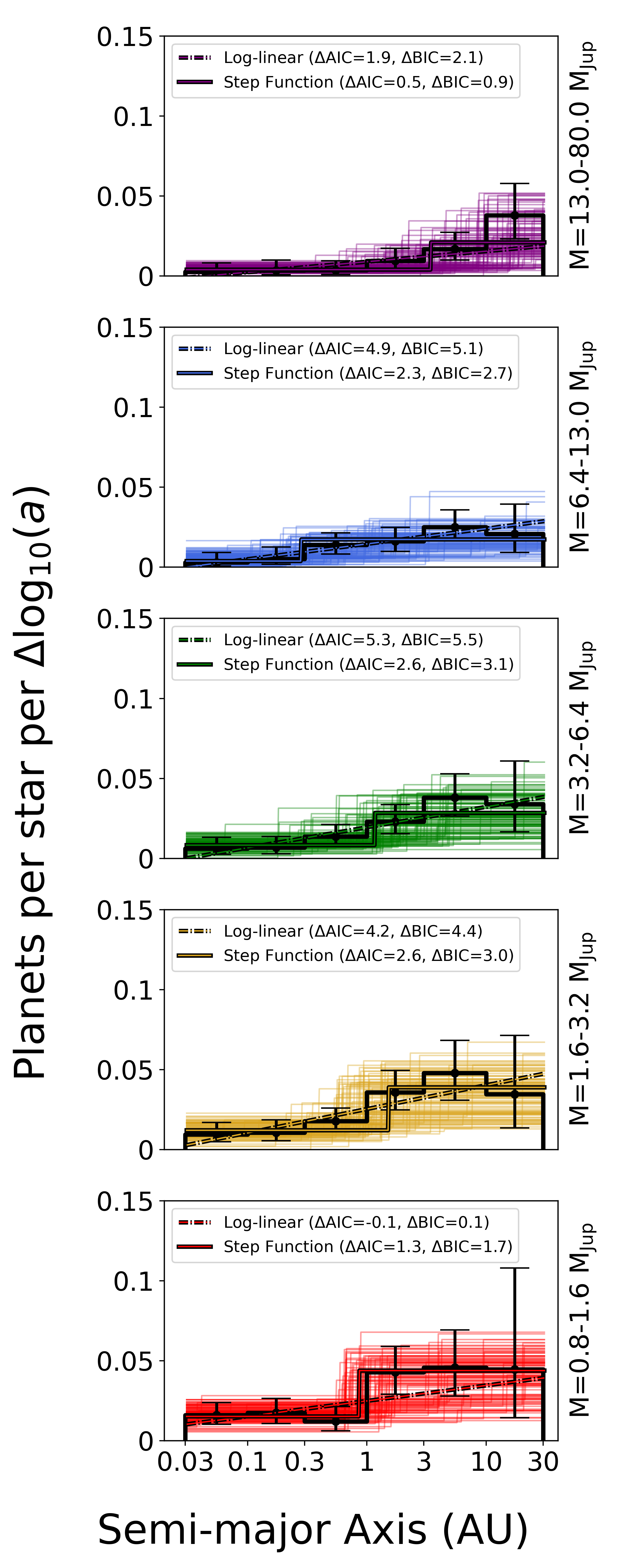}
    \caption{Radial occurrence rate density distributions for companions between 0.8--80 \Mjup, with the highest-mass interval shown in the top panel. We fit each distribution with a log-linear model and a two-level step function. We calculated $\Delta$AIC and $\Delta$BIC values for the best-fit parameters for each model, using a constant model for comparison. In all cases we found that neither model was strongly preferred, though the step function was marginally preferred for oompanions with $M_c$=0.8--1.6 \Mjup, a possible indication that the abundance of planetesimals near the ice line more strongly influences planet formation in this mass interval.}
    \label{fig:mass_sma_occurrence}
\end{figure}

We measured occurrence as a function of both mass and separation in 30 cells defined by the following boundaries: $\frac{M_c}{\Mjup}=$\{0.8, 1.6, 3.2, 6.4, 13, 80\}, $\frac{a}{AU}=$\{0.03, 0.1, 0.3, 1, 3, 10, 30\}. We highlight our region of interest in the context of the CLS true mass catalog in the right panel of Figure \ref{fig:catalog_multipanel}. Figure \ref{fig:mass_sma_occurrence} shows the results of these calculations and we provide the occurrence rates in Table \ref{tab:occurrence_rates}. The drawback of calculating occurrence in 30 separate bin is that each bin contains few companion detections, yielding large uncertainties on our measurements. We therefore performed a simple model comparison for the semi-major axis distribution of each mass bin: a step function versus a log-linear model, given respectively by

\begin{equation}
    \lambda_a(a|a_0, C_1, C_2) =
    \begin{cases}
    C_1 & \text{if } a \leq a_0,\\
    C_2  & \text{otherwise}
    \end{cases}
    \label{eq:step_function}
\end{equation}

\noindent and 

\begin{equation}
    \lambda_a(a|m, b) =
    m \cdot \ln(a) + b,
    \label{eq:log_linear}
\end{equation}

\noindent where $\lambda_a$ is a one-dimensional ORD, that is, the number of planets per star per semi-major axis interval.

We quantified the performance of each model by calculating the difference in Akaike Information Criterion ($\Delta$AIC; \citealt{Akaike1974}) and Bayesian Information Criterion ($\Delta$BIC; \citealt{Schwarz1978}) between the best-fit model and a constant (i.e., flat line) model. In all except the lowest-mass bin, we found a marginal preference for the log-linear model over the step function, indicating a gradual increase in occurrence at wider separations. In the lowest-mass bin, the preference for a step function suggests a more abrupt jump in occurrence near 1 AU. This preference, though modest in significance, is consistent with the expectation that the formation of Jovian planets through pebble accretion would benefit from the increase in planetesimals near the ice line \citep{DrazkowskaAlibert2017, Hyodo2019}.

\subsection{The Brown Dwarf Desert at 1--10 AU}
\label{subsec:mass_occurrence}

\begin{figure}
    \centering
    \includegraphics[width=1.0\linewidth]{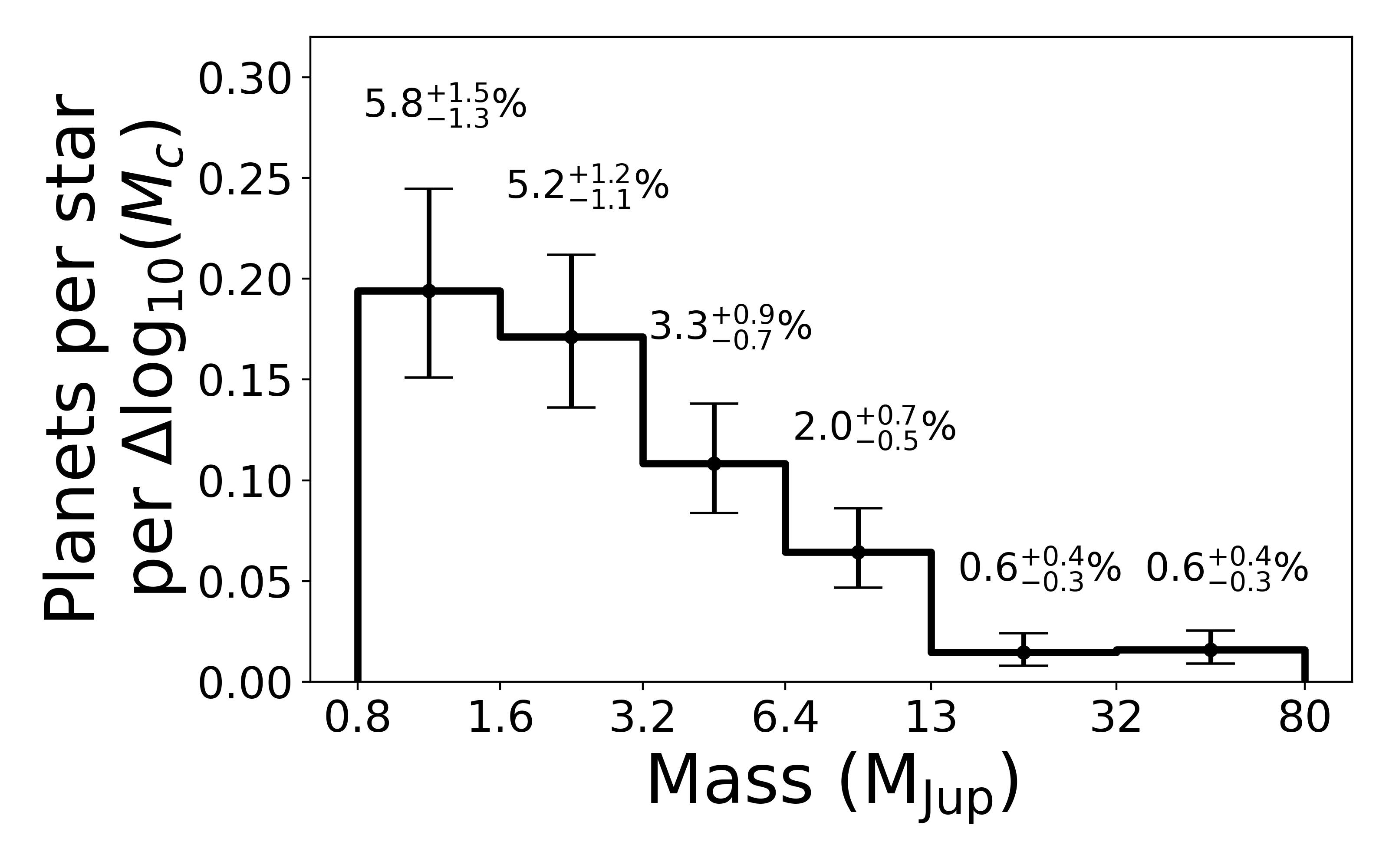}
    \caption{Marginalized occurrence rate density of CLS companions between 1--10 AU. Companion occurrence decreases log-linearly with increasing mass up to $\sim$20 \Mjup, disfavoring a `cut-off' mass at which companion formation near the ice line abruptly becomes inefficient. We annotate each mass bin with the true occurrence rate. The total occurrence rate between 13--80 $\Mjup$ is $1.1^{+0.5}_{-0.4}\%$.}
    \label{fig:mass_occurrence}
\end{figure}

One of the primary strengths of the CLS is its sensitivity to intermediate-separation companions ($\lessapprox$10 AU), uniting the outer limit of separations reached by RV surveys to the inner limit of those probed by direct imaging surveys \citep{BowlerNielsen2018}. We measured the occurrence of giant companions ($0.8 < \frac{M_c}{\Mjup} < 80$) as a function of mass only, marginalizing over the separation interval 1--10 AU and using the same mass bins as in Section \ref{subsec:mass_sma_occurrence}, with the exception of the brown dwarf bin (13--80 \Mjup), which we divided at 32 \Mjup. We found that the occurrence rate decreases monotonically with increasing mass, and that this trend persists in the occurrence rate density, which normalizes the occurrence rate in each mass interval by the interval size. We show the occurrence rate density distribution in Figure \ref{fig:mass_occurrence} and list the corresponding occurrence rates in Table \ref{tab:occurrence_rates}. The occurrence reaches a minimum for companions between 13--80 \Mjup, with a total rate of $1.1^{+0.5}_{-0.4}\%$ in this range. The paucity of brown dwarf companions in this separation range demonstrates that the brown dwarf desert \citep{GretherLineweaver2006} extends to at least 10 AU.

\subsection{A Rise in Giant Companion Occurrence at the Ice Line}
\label{subsec:sma_occurrence}

\begin{figure}[htbp]
\includegraphics[width=0.49\textwidth]{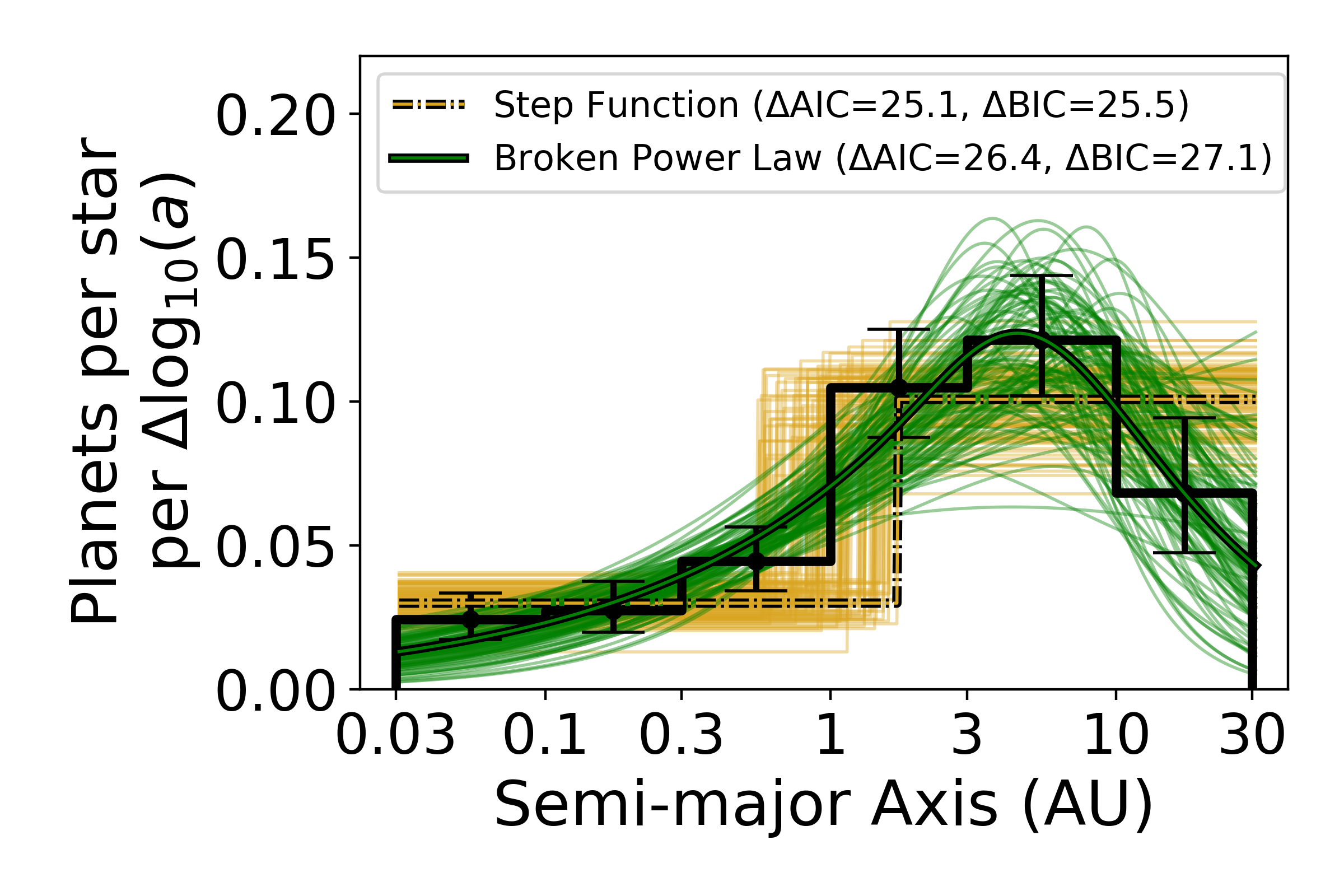}
    \caption{Occurrence rate density between 0.03-30 AU for companions between 0.8--80 \Mjup. Companion occurrence is enhanced beyond 1 AU, and consistent with a decrease beyond 10 AU. A broken power law (solid green) and two-level step function (dotted yellow) fit the distributions comparably well, as measured by their $\Delta$AIC and $\Delta$BIC statistics compared to a constant occurrence rate model. Faded lines show posterior draws for each model. Nearly all steps occur between 0.5 and 1.7 AU.}
    \label{fig:model_comparison}
\end{figure}

Finally, we examined the occurrence distribution of giant companions as a function of semi-major axis. We used the same separation bins as in Section \ref{subsec:mass_sma_occurrence}, this time marginalizing over companion masses between 0.8--80 \Mjup. We found that companion occurrence is approximately constant at close separations, increases abruptly near 1 AU, and remains enhanced to around 10 AU. There is also a marginally significant decrease in occurrence beyond 10 AU, though the uncertainty on our occurrence measurement between 10--30 AU prevents a firm conclusion. Figure \ref{fig:model_comparison} shows our marginalized semi-major axis distribution, with the corresponding occurrence rates given in Table \ref{tab:occurrence_rates}. Comparing this distribution with those we calculated for separate mass intervals (see Section 
\ref{subsec:mass_sma_occurrence}), we see that the relatively high occurrence between 1--10 AU exhibited in the low-mass bins contributes to the rise at 1 AU in the marginalized distribution. Meanwhile, the small number of detections between 10--30 AU in all mass intervals results in comparatively low occurrence in this range. We fit this distribution with both a two-level step function and a broken power law. Our broken power law model matches that of \cite{Fulton2021} and is given by

\begin{equation}
    \lambda_a(a|C, \beta, a_0, \gamma) = C \left ( \frac{a}{\text{AU}} \right )^{\beta} \left (1-e^{-(\frac{a}{a_0})^{\gamma}} \right ),
    \label{eq:broken_power_law}
\end{equation}

\noindent where C is a normalization constant, $\beta$ is the power law index at separations beyond the ``break'' separation $a_0$, and $\beta+\gamma$ is the power law index interior to $a_0$. We found that both models performed comparably well, and that the step location reliably fell near 1 AU.

We repeated these calculations using companion minimum mass and modified parameter bounds to match the analysis of \cite{Fulton2021}, who calculated the semi-major axis distribution between 0.1--30 AU for companions with \Msini=30--6000 $M_{\oplus}$ (0.09--18.9 \Mjup). The maximum likelihood  parameters of our broken power law fit to this occurrence distribution agreed closely with those derived by \cite{Fulton2021}, and we recovered the two key features noted in that study: an abrupt rise in occurrence near 1 AU and a possible fall-off beyond 10 AU. The persistence of the increase in occurrence near 1 AU through changes in binning, mass limits, and $M_c$ versus \Msini suggests that it is a robust feature of the radial occurrence distribution.

\section{Discussion}
\label{sec:discussion}

\subsection{An Astrophysically Motivated Planet--Brown Dwarf Divide}
\label{subsec:discussion_planet_bd_divide}

The International Astronomical Union defined brown dwarfs in 2003 as ``substellar objects with true masses above the limiting mass for thermonuclear fusion of deuterium [$\sim$13 \Mjup] ... no matter how they formed nor where they are located."\footnote{\url{https://web.archive.org/web/20141216075559/http://home.dtm.ciw.edu/users/boss/definition.html}} This categorization had already been questioned by that time \citep{Burrows2001}, and has faced further scrutiny in the years since (e.g., \citealt{Spiegel2011, Schlaufman2018}). Despite the benefit of providing an objective dividing line based on the measurable quantity $M_c$, there is no evidence that the processes that form giant planets and brown dwarfs are directly tied to the deuterium burning limit. To better address the nuances of companion formation at the planet--brown dwarf transition, astronomers have searched for definitions motivated by factors such as formation pathway, dynamics, and host star properties. \cite{Schlaufman2018} found evidence for a divide between objects above/below 4--10 \Mjup based on the preference for objects below this range to orbit metal-rich host stars, a trait not exhibited by objects above it. Combining high-contrast imaging and astrometric observations between 5--100 AU, \cite{Bowler2020} found that companions with $M_c<15$ \Mjup have preferentially low eccentricities, peaking at $e\lesssim0.2$, whereas higher-mass companions exhibit a broad range of eccentricities between $e$=0.6--0.9. Using a subset of the companion sample of \cite{Bowler2020} but a different framework for calculating parameter priors, \cite{DoO2023} found insufficient evidence for a distinction in the eccentricity distributions between low- and high-mass companions. Recent work by our team \citep{Giacalone2025, Gilbert2026} examines the role of both metallicity and eccentricity in the planet--BD distinction.

We searched for a similar distinction in the radial occurrence distributions of planets and BDs in Section \ref{subsec:mass_sma_occurrence}. Among the five mass bins we analyzed, we did not find any abrupt change in the radial distributions with increasing mass. The lack of a sharp dividing line in mass among these distributions suggests that ``bottom-up" and ``top-down" formation mechanisms may produce companions in overlapping mass ranges. Because the majority of our cells contained fewer than five objects in this experiment, we were unable to constrain occurrence at high significance. Thus, some latent separation-dependent trends may exist that will provide a clearer dividing line between planets and brown dwarfs. On the other hand, we also found smooth variation with increasing mass for the marginalized occurrence distribution in Section \ref{subsec:mass_occurrence}, supporting the idea of a ``fuzzy" planet--BD boundary.

We found marginal evidence for jumps in occurrence near 1 AU for companions between 0.8--3.2 \Mjup. These features likely drive the analogous and more significant jump in the marginalized radial distribution (Section \ref{subsec:sma_occurrence}). Increased core accretion efficiency is expected near the ice line due to the enhancement of icy pebbles and silicate dust that can coagulate via gravitational and/or streaming instabilities in this region \citep{DrazkowskaAlibert2017, Hyodo2019}. Thus, the presence or absence of an abrupt increase in occurrence near the ice line may be used to infer whether an object formed by core accretion or gas/disk instability. Increased Doppler baselines, larger stellar samples, and superior orbital fits through the inclusion of Gaia DR4 astrometry \citep{Winn2022} will enable occurrence measurements precise enough to confirm or rule out an occurrence jump in narrow mass intervals such as those examined here. Such measurements will place an astrophysically motivated planet--brown dwarf dividing line within reach. 

\subsection{The Brown Dwarf Desert}
\label{subsec:discussion_bd_desert}

We showed in Section \ref{subsec:mass_occurrence} that occurrence decreases with increasing mass, reaching a minimum for brown dwarfs (13--80 \Mjup). This interval extends to a mass ratio ($q=M_c/M_{\text{pri}}$) of approximately $q=0.1$ for the Sun-like stars that comprise most of the CLS. Although we were unable to measure occurrence rates for higher-mass (stellar) companions, studies of stellar multiplicity over wide ranges in separation and primary type have found that the occurrence rate increases with increasing mass ratio for low $q$ ($\sim$0.1--0.3; \citealt{Reggiani2013, Moe2017}). Assuming this pattern holds for solar hosts between 1--10 AU, the 13--80 \Mjup interval represents a valley in the companion occurrence distribution, meaning that the brown dwarf desert found by \cite{GretherLineweaver2006} for separations $\lesssim$3 AU reaches, and perhaps exceeds, 10 AU. This feature was observed recently by \cite{An2025}, who acknowledged that their sample was not designed for occurrence rate calculations. Its presence in both samples provides strong evidence that the brown dwarf desert reaches 10 AU.

The extension of the brown dwarf desert to at least 10 AU has ramifications for theories of brown dwarf dynamical evolution. In addition to a brown dwarf desert out to 5-year orbital periods ($\sim$3 AU), \cite{GretherLineweaver2006} reported a lack of such a desert among free-floating BDs in the Orion, Pleiades, and M35 stellar clusters. In other words, brown dwarfs often form in isolation, but rarely as companions to Sun-like stars. They argued that this distinction disfavors a minimum in the efficiency of gravitational collapse or disk fragmentation \citep{IdaLin2004-metallicity} as the source of the BD desert, and instead supports post-formation migration as the mechanism preventing the settlement of BDs at close separations. Our findings are consistent with this picture, but require that such migration mechanisms be capable of explaining the BD desert out to 10 AU.

\subsection{Comparison to Direct Imaging Surveys}
\label{subsec:discussion_imaging}

The CLS's achievement of sensitivity to companions near 10 AU is an important bridge between the short- and long-period regimes so far probed by RV and imaging surveys, respectively. It has allowed for comparison between occurrence rates calculated by these two methods \citep{Fulton2021}, under the assumption that giant planet occurrence remains constant between young ($\sim$10 Myr) and intermediate (Gyr) stellar ages. 

In this work, we addressed a remaining barrier to direct comparison, which is the discrepancy between the minimum masses measured by RVs alone and the true masses estimated from direct imaging through companion formation and luminosity models \citep{Baraffe03, Nielsen2019}. We measured the occurrence of companions between 0.03--30 AU and 0.8--80 \Mjup, both as functions of mass and separation together (Section \ref{subsec:mass_sma_occurrence}) and of each parameter separately (Sections \ref{subsec:mass_occurrence} and \ref{subsec:sma_occurrence}).

In Section \ref{subsec:mass_occurrence}, we found an occurrence rate of $1.1^{+0.5}_{-0.4}\%$ for brown dwarfs ($M_c$=13--80 \Mjup) between 1--10 AU. Assuming the occurrence rate is log-constant at wide separations, this value is in close agreement with that of \cite{Nielsen2019}, who estimated a BD occurrence rate of $0.8^{+0.8}_{-0.5}\%$ between 10--100 AU from observations made with the Gemini Planet Imager \citep{Macintosh2014}. If we extrapolate our BD occurrence rate between 10--30~AU ($1.6^{+0.9}_{-0.6}\%$) to 10--100~AU, assuming a log-uniform rate, we arrive at $3.2^{+1.3}_{-0.8}$\%, larger than the estimate of \cite{Nielsen2019}. We note a number of caveats in our comparison: the large fractional uncertainties in both measurements, the difference in primary star masses and ages in the two samples, and the uncertainty in the period distribution of brown dwarfs between 10--100~AU.

\subsection{Distribution of \texttt{Orvara} Inclination Draws}
\label{subsec:discussion_inclination_distributions}

\begin{figure*}[t]
    \centering
    \begin{minipage}[b]{0.49\linewidth}
        \centering
        \includegraphics[width=\linewidth]{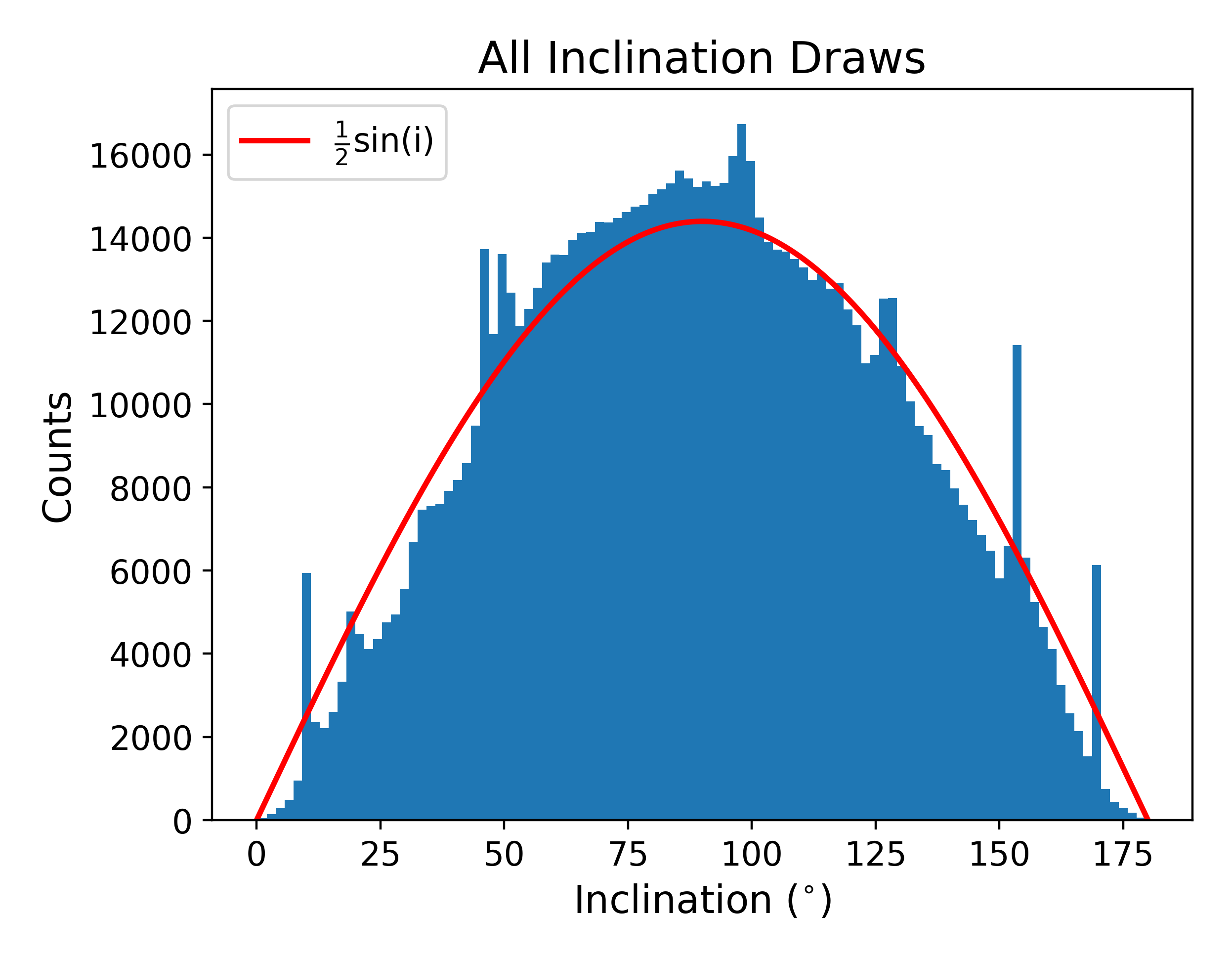}
    \end{minipage}
    \hfill
    \begin{minipage}[b]{0.49\linewidth}
        \centering
        \includegraphics[width=\linewidth]{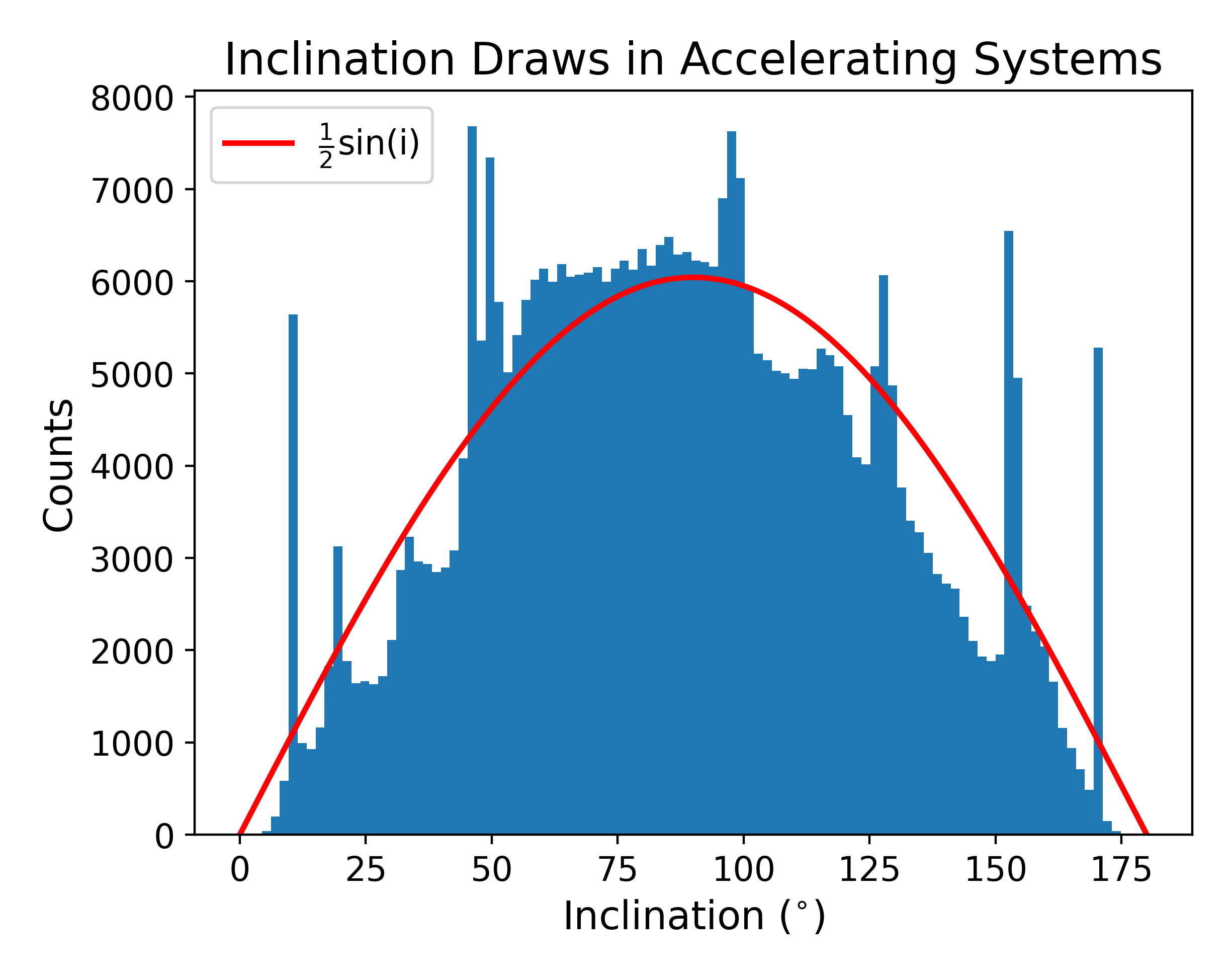}
    \end{minipage}
    
    \caption{\textbf{Left:} Posterior distribution of inclinations from all \texttt{Orvara} chains in our sample, including all companions in multi-companion systems. The red line shows a sinusoidal distribution, which a large sample of randomly-oriented companion inclinations is expected to follow. We scaled this distribution so that its integral equals that of the posterior. \textbf{Right:} The same as the right panel, but for only the 53 systems in which we calculated a significant ($\geq 3\sigma$) astrometric acceleration. As expected, we find greater deviation from the theoretical distribution among this sample, which is biased to include massive companions on inclined orbits. Nevertheless, the roughly sinusoidal distribution of even these inclinations suggests that our fitting procedure did not introduce heavy bias to our analysis.}
    \label{fig:inclination_distribution}
\end{figure*}

Following \cite{An2025}, we validated our \texttt{Orvara} fits by evaluating the distribution of posterior orbital inclination draws among all companion fits. Assuming an unbiased sample, we expect the distribution of sampled inclinations to match the geometric inclination prior, that is, the distribution that arises from an ensemble of orbits randomly oriented in three dimensions. A deviation from this prior could indicate systematic biases in our sample selection or orbit fitting procedures. After combining all MCMC chains across all fitted companions, we find that the inclination distribution fits a sinusoid well. Additionally, we examined the inclination distribution of the companions in the 53 systems in our sample exhibiting a significant HGCA acceleration ($\Delta \mu/\sigma_{\Delta \mu}\geq 3$). We found that this distribution was likewise sinusoidal, though with greater deviation as compared to the full sample. This deviation is expected, given that filtering on astrometric acceleration significance introduces a preference for massive companions on inclined orbits. Figure \ref{fig:inclination_distribution} depicts the inclination distributions of our full and astrometrically accelerating samples.

\subsection{Biases}
\label{subsec:discussion_biases}

Our study is subject to multiple biases, which we list here. First, as noted in Section \ref{subsec:discussion_inclination_distributions}, we incorporated HGCA data into our fits only for systems exhibiting significant astrometric accelerations. Because larger companions produce more significant accelerations, our strategy allowed us to identify massive inclined companions, but was unable to precisely constrain the inclinations of their low-mass counterparts. This is analogous to the RV bias against low-mass, wide-separation objects, which we accounted for using a system-by-system sensitivity correction. Additionally, as noted by \cite{An2025}, the RV bias toward edge-on orbits and the astrometric bias toward face-on orbits implies that a combination of these two techniques can mitigate the overall inclination bias. Nevertheless, a detailed accounting of the HGCA's sensitivity would improve our result by accounting for the small inclined objects that the HGCA missed.

Another limitation in our analysis is the exclusion of RV trends. Long-period companions produce non-periodic signals that are consistent with a wide range of companion models, often including planets, brown dwarfs, and stars. \cite{Rosenthal2021} reported trends in 89 systems, nearly as many as the number of systems hosting companions with full orbits. Although it is possible to model such signals probabilistically (e.g. \citealt{VanZandt2024, VanZandt2025}), we opted not to include them in our analysis due to their comparatively minor contribution to occurrence statistics in our parameter space of interest. A future study incorporating trends would strengthen occurrence constraints, particularly at wide separations.

\section{Conclusion}
\label{sec:conclusion}
We re-fit the orbits of 191 substellar and 3 stellar companions from the California Legacy Survey to measure occurrence at wide separations ($\lesssim$30 AU). We combined the multi-decade RV data sets of the California Legacy Survey with absolute astrometry from the $Hipparcos$-$Gaia$ Catalog of Accelerations to fit three-dimensional orbits where possible. We identified seven of the 18 apparently-substellar companions ($M_c \sin i$ = 13--80 \Mjup) in the CLS catalog as inclined M dwarfs.

We modeled companion occurrence as a Poisson point process in independent rectangular cells in $\log M- \log a$ space, accounting for survey sensitivity and uncertainties on fitted orbital parameters. We examined planet and BD occurrence as a function of both mass and separation. Our main conclusions are as follows:

\begin{itemize}
    \item Nearly 50\% of the RV brown dwarfs we refit were in fact low-mass stars, illustrating that populations of intrinsically rare objects are particularly susceptible to contamination from higher-mass companions on inclined orbits.

    \item The lack of a sharp transition between either the shape of the semi-major axis distributions of objects between 0.8--80 \Mjup (Section \ref{subsec:mass_sma_occurrence}), or the marginalized occurrence rate between 1--10 AU (Section \ref{subsec:mass_occurrence}), disfavors a ``cut-off mass" below which planets are exclusively formed by core accretion and above which brown dwarfs are exclusively formed by gravitational instability. Instead, the smooth variation in both distributions suggests that the two mechanisms produce companions in overlapping mass ranges.
    
    \item Companion occurrence between 1--10 AU is a monotonically decreasing function of mass, reaching a minimum for brown dwarfs (13--80 \Mjup), consistent with previous findings of a ``brown dwarf desert" at shorter separations. Whatever mechanism(s) stifle the formation/migration of brown dwarfs at low separations are relevant at wider separations as well.

    \item The radial occurrence distribution of all companions between 0.8--80 \Mjup exhibits a significant preference for a step near 1 AU, consistent with \cite{Fulton2021}'s finding of an enhancement in the occurrence of companions with $M \sin i$= 30--6000 $M_{\oplus}$ beyond 1 AU. Determining precisely which mass sub-intervals drive this feature will help determine the dividing line between companions generated by core accretion versus gravitational/disk instability. The possibility of a drop-off in occurrence beyond 10 AU persists as well, though the large uncertainties at these wide separations prevent a definitive conclusion.
\end{itemize}

The combination of RVs and absolute astrometry is a powerful tool for the detection and characterization of exoplanets. By breaking the RV mass-inclination degeneracy, this technique facilitates direct comparison of dynamical and model masses across different detection methods and semi-major axis ranges. The upcoming release of Gaia DR4, expected December 2026,\footnote{\url{https://www.cosmos.esa.int/web/gaia/release}} will both increase the available pool of giant companion-hosting stars and allow for even more precise orbital modeling, sharpening our understanding of the trends described in this work.

\appendix

\FloatBarrier
\section{Occurrence Rate Table}
\label{app:occurrence_rate_tables}
\begin{table*}[t]
\centering
\caption{Derived Occurrence Rates}
\label{tab:occurrence_rates}

\begin{minipage}[t]{0.48\textwidth}
\centering
\begin{tabular}{ccc}
\hline
$a$ & $M_c$ & Occurrence Rate \\
(AU) & ($\Mjup$) & (Companions per star) \\
\hline
0.03--0.1 & 0.8--1.6 & $0.008^{+0.004}_{-0.003}$ \\
 0.1--0.3 & 0.8--1.6 & $0.008^{+0.004}_{-0.003}$ \\
   0.3--1 & 0.8--1.6 & $0.006^{+0.005}_{-0.003}$ \\
     1--3 & 0.8--1.6 & $0.020^{+0.008}_{-0.007}$ \\
    3--10 & 0.8--1.6 & $0.024^{+0.012}_{-0.009}$ \\
   10--30 & 0.8--1.6 &    $0.02^{+0.03}_{-0.01}$ \\
0.03--0.1 & 1.6--3.2 & $0.005^{+0.004}_{-0.002}$ \\
 0.1--0.3 & 1.6--3.2 & $0.005^{+0.004}_{-0.002}$ \\
   0.3--1 & 1.6--3.2 & $0.009^{+0.004}_{-0.003}$ \\
     1--3 & 1.6--3.2 & $0.017^{+0.007}_{-0.005}$ \\
    3--10 & 1.6--3.2 & $0.025^{+0.011}_{-0.009}$ \\
   10--30 & 1.6--3.2 & $0.016^{+0.018}_{-0.010}$ \\
0.03--0.1 & 3.2--6.4 & $0.003^{+0.004}_{-0.002}$ \\
 0.1--0.3 & 3.2--6.4 & $0.003^{+0.003}_{-0.002}$ \\
   0.3--1 & 3.2--6.4 & $0.007^{+0.004}_{-0.003}$ \\
     1--3 & 3.2--6.4 & $0.011^{+0.005}_{-0.004}$ \\
    3--10 & 3.2--6.4 & $0.020^{+0.008}_{-0.006}$ \\
   10--30 & 3.2--6.4 & $0.016^{+0.013}_{-0.008}$ \\
0.03--0.1 &  6.4--13 & $0.002^{+0.003}_{-0.001}$ \\
 0.1--0.3 &  6.4--13 & $0.002^{+0.004}_{-0.002}$ \\
   0.3--1 &  6.4--13 & $0.007^{+0.004}_{-0.003}$ \\
     1--3 &  6.4--13 & $0.008^{+0.004}_{-0.003}$ \\
    3--10 &  6.4--13 & $0.013^{+0.006}_{-0.004}$ \\
   10--30 &  6.4--13 & $0.010^{+0.009}_{-0.006}$ \\
0.03--0.1 &   13--80 & $0.001^{+0.003}_{-0.001}$ \\
 0.1--0.3 &   13--80 & $0.002^{+0.003}_{-0.001}$ \\
   0.3--1 &   13--80 & $0.002^{+0.003}_{-0.001}$ \\
     1--3 &   13--80 & $0.005^{+0.004}_{-0.002}$ \\
    3--10 &   13--80 & $0.009^{+0.005}_{-0.004}$ \\
   10--30 &   13--80 & $0.018^{+0.009}_{-0.007}$ \\
\end{tabular}
\end{minipage}
\hfill
\begin{minipage}[t]{0.48\textwidth}
\centering
\begin{tabular}{ccc}
\hline
$a$ & $M_c$ & Occurrence Rate \\
(AU) & ($\Mjup$) & (Companions per star) \\
\hline
 1--10 & 0.8--1.6 &    $0.06^{+0.02}_{-0.01}$ \\
 1--10 & 1.6--3.2 &    $0.05^{+0.01}_{-0.01}$ \\
 1--10 & 3.2--6.4 & $0.033^{+0.009}_{-0.007}$ \\
 1--10 &  6.4--13 & $0.020^{+0.007}_{-0.005}$ \\
 1--10 &   13--32 & $0.006^{+0.004}_{-0.003}$ \\
 1--10 &   32--80 & $0.006^{+0.004}_{-0.003}$ \\
0.03--0.1 & 0.8--80 & $0.013^{+0.005}_{-0.004}$ \\
 0.1--0.3 & 0.8--80 & $0.013^{+0.005}_{-0.004}$ \\
   0.3--1 & 0.8--80 & $0.023^{+0.006}_{-0.005}$ \\
     1--3 & 0.8--80 & $0.050^{+0.010}_{-0.008}$ \\
    3--10 & 0.8--80 &    $0.06^{+0.01}_{-0.01}$ \\
   10--30 & 0.8--80 &    $0.03^{+0.01}_{-0.01}$ \\
\end{tabular}
\end{minipage}

\begin{tablenotes}
\footnotesize
\item Note. The left table gives occurrence rates corresponding to the histograms in Figure \ref{fig:mass_sma_occurrence}. The right column gives rates for the histograms in Figures \ref{fig:mass_occurrence} and \ref{fig:model_comparison}. Measurements are given as posterior medians with 16th and 84th percentile ranges.
\end{tablenotes}
\end{table*}

\FloatBarrier
\section{Highlighted Outliers in Refit Catalog}
\label{app:highlighted_outliers}

\begin{figure*}[t]

        \centering
        \includegraphics[width=\linewidth]{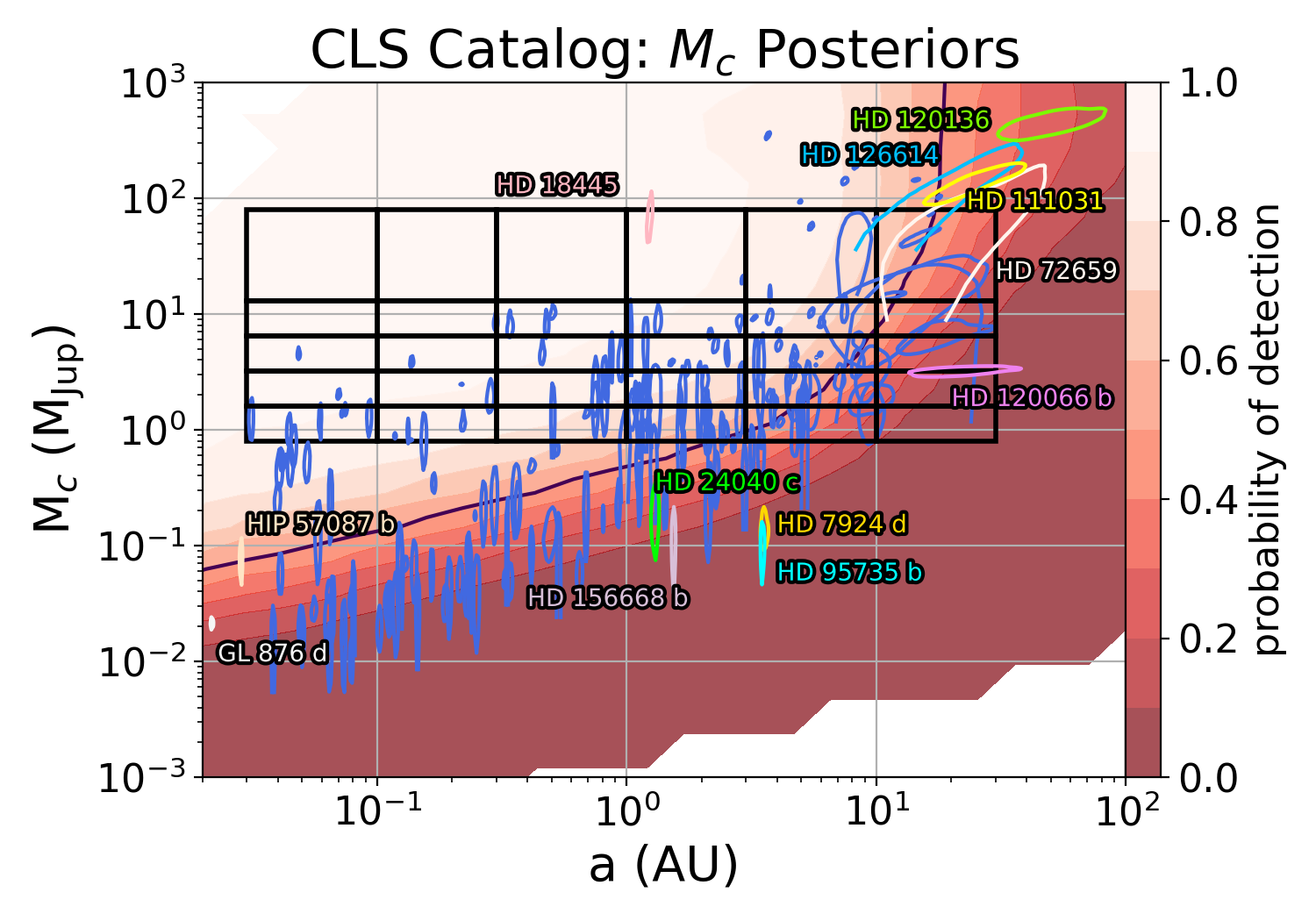}
    
    \caption{Same as Figure \ref{fig:catalog_multipanel} with a handful of companions in the extremities of the parameter space identified with separate colors and labels. Companions such as HD 7924 d and HD 95735 b highlight an important caveat to our analysis: although the survey-averaged sensitivity suggests that these planets would not be detectable, CLS achieved higher sensitivity to such objects in their host systems. The companion to HD 18445 straddles the 80 $\Mjup$ boundary, and our refit did little to constrain its mass due to its high eccentricity and sparse RV sampling. Although this object could be a low-mass M dwarf, we did not count it among the seven spurious RV brown dwarfs in the CLS catalog.}
    \label{fig:catalog_appendix}
\end{figure*}

\FloatBarrier
\section{Original and Refit Companion Parameters}
\label{app:all_before_after_table}

The table below provides orbital parameters for the 194 CLS companions we fit in our analysis. The left-most column shows the companion name as it appears in Table 3 or 4 of \cite{Rosenthal2021}. Subscripts $i$ and $f$ respectively indicate parameters reported by \cite{Rosenthal2021} and by this work, with the exception of HD 192310, HD 219134, HD 28185, and GL 876, for which we adopted literature values as described in Section \ref{sec:analysis}. For these systems, blank fields indicate that we adopted the values given by \cite{Rosenthal2021}, if available, for our occurrence analysis.

\begin{longrotatetable}
\begin{deluxetable*}{cccccccccccc}
\tablecaption{Original and Refit Companion Parameters}
\label{tab:all_before_after}
\tabletypesize{\scriptsize}

\tablehead{
CLS & $a_i$ & $M_c \sin i$ & $e_i$ & $a_f$ & $M_c$ & $e_f$ &
$\omega_f$ & $i_f$ & $\Omega_f$ & $t_{p,f}$ & HGCA? \\
Name & (AU) & ($M_{\rm Jup}$) &  & (AU) & ($M_{\rm Jup}$) &  &
(deg) & (deg) & (deg) & (JD)
}
\startdata
 BD-103166 b & $0.0450^{+0.0006}_{-0.0007}$ &       $0.45^{+0.02}_{-0.02}$ &       $0.02^{+0.02}_{-0.02}$ &    $0.0450^{+0.0006}_{-0.0006}$ &    $0.51^{+0.24}_{-0.06}$ &       $0.02^{+0.02}_{-0.01}$ &     $-0^{+200}_{-100}$ &         $90^{+40}_{-40}$ &  $200^{+100}_{-100}$ &       $2455198.4^{+0.9}_{-1.0}$ &       False \\
    GL 317 b &    $1.180^{+0.008}_{-0.008}$ &       $1.85^{+0.04}_{-0.04}$ &       $0.10^{+0.02}_{-0.02}$ &       $1.182^{+0.008}_{-0.008}$ &       $2.1^{+0.8}_{-0.2}$ &       $0.11^{+0.02}_{-0.02}$ &      $-20^{+10}_{-10}$ &         $90^{+40}_{-40}$ &  $200^{+100}_{-100}$ &           $2455800^{+30}_{-30}$ &       False \\
    GL 317 c &          $5.8^{+0.7}_{-0.4}$ &       $1.67^{+0.08}_{-0.08}$ &       $0.25^{+0.11}_{-0.07}$ &             $5.5^{+0.3}_{-0.2}$ &       $1.9^{+1.1}_{-0.3}$ &       $0.19^{+0.07}_{-0.06}$ &      $130^{+20}_{-20}$ &         $90^{+40}_{-40}$ &  $200^{+100}_{-100}$ &         $2456900^{+300}_{-300}$ &       False \\
    GL 687 b &    $0.166^{+0.001}_{-0.001}$ &    $0.055^{+0.005}_{-0.005}$ &       $0.15^{+0.08}_{-0.08}$ &       $0.166^{+0.001}_{-0.001}$ & $0.063^{+0.035}_{-0.009}$ &       $0.08^{+0.07}_{-0.05}$ &       $50^{+60}_{-90}$ &         $90^{+40}_{-40}$ &  $200^{+100}_{-100}$ &            $2455206^{+17}_{-6}$ &       False \\
    GL 876 d & $0.0218^{+0.0002}_{-0.0002}$ &    $0.018^{+0.002}_{-0.002}$ &       $0.27^{+0.09}_{-0.10}$ &                             --- & $0.021^{+0.002}_{-0.002}$ &                          --- &                    --- &                   $59.5$ &                  --- &                             --- &       False \\
    GL 876 c &    $0.136^{+0.001}_{-0.001}$ &       $0.70^{+0.01}_{-0.01}$ &    $0.006^{+0.006}_{-0.004}$ &                             --- &    $0.81^{+0.02}_{-0.02}$ &                          --- &                    --- &                   $59.5$ &                  --- &                             --- &       False \\
\enddata
\end{deluxetable*}
\end{longrotatetable}


\software{\texttt{astropy} \citep{astropy:2018},
          \texttt{Orvara} \citep{Orvara2021}, \texttt{numpy} \citep{numpy:2020}, 
          \texttt{scipy} \citep{scipy:2020}
          }

\bibliography{sj-bd}
\bibliographystyle{aasjournal}

\end{document}